# $\beta$-Ga$_2$O$_3$ MOSFETs with near 50 GHz f$_{MAX}$ and 5.4 MV/cm average breakdown field

Chinmoy Nath Saha, Abhishek Vaidya, A F M Anhar Uddin Bhuiyan, Lingyu Meng, Hongping Zhao, *Senior Member, IEEE* and Uttam Singisetti *Senior Member, IEEE*

***Abstract*— This letter reports high-performance $\beta$-Ga$_2$O$_3$ thin channel MOSFETs with T-gate and degenerately doped source/drain contacts regrown by MOCVD. Gate length scaling (L$_G$= 160-200 nm) leads to a peak drain current (I$_{D,MAX}$) of 285 mA/mm and peak trans-conductance (g$_m$) of 52 mS/mm at 10 V drain bias with 23.5 $\Omega$ mm on resistance (R$_{on}$). A low metal/n+ contact resistance of 0.078 $\Omega$ mm was extracted from TLM measurement. R$_{on}$ is dominated by interface resistance between channel and regrown layer. A gate-to-drain breakdown voltage of 192 V is measured for L$_{GD}$ = 355 nm resulting in average breakdown field (E$_{AVG}$) of 5.4 MV/cm. This E$_{AVG}$ is the highest reported among all sub-micron gate length lateral FETs. RF measurements on 200 nm Silicon Nitride (Si$_3$N$_4$) passivated device shows a current gain cut off frequency (f$_T$) of 11 GHz and record power gain cut off frequency (f$_{MAX}$) of 48 GHz. The f$_T$.V$_{Br}$ product is 2.11 THz.V for 192 V breakdown voltage. The switching figure of merit exceeds that of silicon and is comparable to mature wide-band gap devices.**

***Index Terms*— Ga$_2$O$_3$, power gain cut off frequency, switching figure of merit.**

## I. Introduction

$\beta$-Ga$_2$O$_3$ has attracted intense attention worldwide because of its favourable material properties [1] for power, RF switching and RF applications [1], [2]. Multi-kV drain breakdown voltages have been reported in FETs [3]–[6] and diodes [7]–[9]. Theoretically calculated large electron saturation velocity [10] also makes Ga$_2$O$_3$ a suitable candidate for next generation RF transistors. Mature bulk crystal growth [11] and thin film growth techniques (MOCVD, MBE, HVPE) with controllable doping (10$^{16}$- 10$^{20}$cm$^{-3}$) [12]–[15] have made device fabrication with novel structures possible including both depletion and enhancement mode FETs. [16]–[19]. Modulation doped $\beta$-(Al$_x$Ga$_{1-x}$)$_2$O$_3$/Ga$_2$O$_3$ Heterostrcuture FET (HFET) have been demonstrated with record 30/37 GHz f$_T$/f$_{MAX}$ [20] and high temperature stability of RF performance up to 250$^0$C [21] Highly doped contact regrowth process using MOCVD and MBE have been reported with very low contact resistance [22], improved g$_m$ and RF performance in MESFETs [23] and AlGaO/GaO HFETs [20], [24]. Thick channel ($\geq$ 200 nm) $\beta$-Ga$_2$O$_3$ MOSFET has been demonstrated but poor gate control because of thicker channel degrades high frequency performances [25]–[27].

In this letter, we report a highly scaled T gate $\beta$-Ga$_2$O$_3$ thin channel MOSFET for achieving better gate control and improved RF performance. We demonstrate devices with simultaneously high I$_{D,max}$ (>250 mA/mm) with low R$_{on}$ (<25 $\Omega$ mm), high average field strength E$_{AVG}$ (>4.5 MV/cm) and f$_T$,f$_{MAX}$ >10 GHz. The device showed a highest f$_{MAX}$ of 48 GHz among Ga$_2$O$_3$ FETs. We demonstrated high average breakdown field without any field plate technique which could potentially provide cost advantages for high voltage RF applications.

## II. Device Structure and Fabrication

The epitaxial structure of the device was grown on (010) Fe-doped semi-insulating Ga$_2$O$_3$ substrate using ozone molecular beam epitaxy (MBE) method follwing conditions described in ref. [28]. The device stack consists of 200 nm unintentionally (UID) doped buffer layer and 60 nm Si doped ( 9.2 x 10$^{17}$ cm$^{-3}$) channel. Device fabrication started with blanket ALD Al$_2$O$_3$, PECVD SiO$_2$ and e-beam evaporated Cr layers. Next, using Cr as a hard mask, SiO$_2$ and Al$_2$O$_3$ layers were removed everywhere except the gate region by reactive ion etching (RIE) and wet etching respectively. The channel was never exposed directly to the RIE. After removal of Cr hard mask using wet etch, a 85 nm thick degenerately doped (1 x 10$^{20}$ cm$^{-3}$) Ga$_2$O$_3$ layer was grown by MOCVD at 700 $^0$C. After regrowth, sample was dipped in buffered HF to remove SiO$_2$/Al$_2$O$_3$ regrowth mask.

Next, device mesa isolation was performed using high power BCl$_3$ based ICP-RIE etch. Ti/Au/Ni metal stack (50 nm/ 120 nm/ 40 nm) was deposited in the source/drain (S/D) regions by electron-beam evaporation. A 20 nm SiO$_2$ was deposited using plasma enhanced ALD for gate dielectric followed by a 450 $^0$C 1 minute annealing under N$_2$ atmosphere to improve the gate oxide quality. SiO$_2$ was removed from S/D contact regions using low power ICP-RIE etch. Finally, Ni/Au metal (50 nm/270 nm) was patterned to form T-shaped gates. A trilayer resist stack (MMA/PMGI/PMMA) [29] was used for fabricating 160-200 nm foot length and 500 nm top gate hat. The device layout was designed for GSG probing with 2x20 $\mu$m width T-layout.

We acknowledge the support from AFOSR under award FA9550-18-1-0479 (Program Manager: Ali Sayir), from NSF under awards ECCS 2019749, ECCS 1919798, from Semiconductor Research Corporation under GRC Task ID 3007.001, and II-VI Foundation Block Gift Program.

Chinmoy Nath Saha, Abhishek Vaidya and Uttam Singisetti are with Electrical Engineering department, University at Buffalo, Buffalo, NY-14260, USA.

A F M Anhar Uddin Bhuiyan, Lingyu Meng, Hongping Zhao are with Department of Electrical and Computer Engineering, The Ohio State University, Columbus, OH 43210, USA

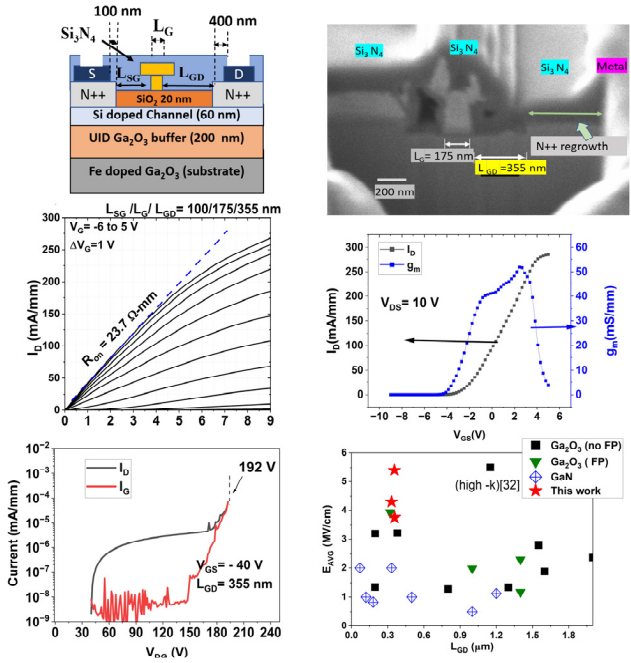

Fig. 1. (a) Cross section schematic of a MOSFET after passivation. (b) FIB-cross section image of a fabricated device after passivation, (c) $I_D$- $V_{DS}$ output curve and (d) $I_D$- $V_{GS}$ Transfer curve of the DUT after passivation showing 285 mA/mm on current at $V_{DS}$ = 10 V (e) Three terminal off-state breakdown measurement for $L_{GD}$= 355 nm showing breakdown ( $V_{DG}$ = $V_{DS}$ - $V_{GS}$) at 192 V. (b) $E_C$ vs $L_{GD}$ reported in literature with and without Field Plate (FP) technique showing that our device has the highest $E_{avg}$ among sub-micron devices

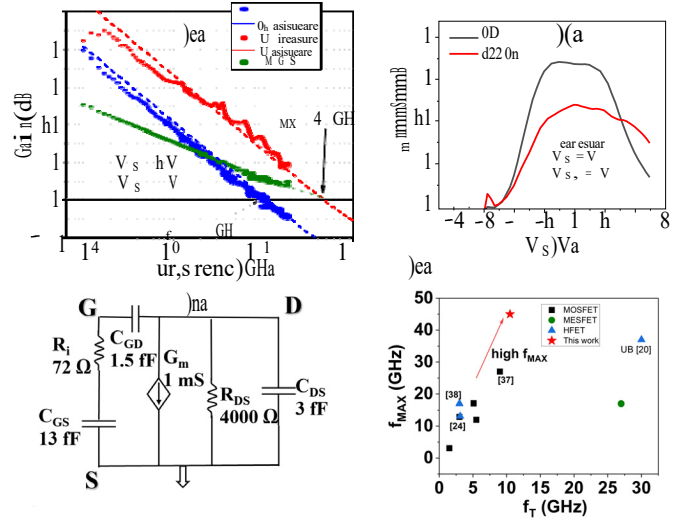

Fig. 2. (a) Measured and simulated small signal RF performance of 175 nm gate length MOSFET showing 48 GHz $f_{MAX}$ (b) small signal equivalent circuit model of our device (c) Pulsed $I_D$-$V_G$ transfer curve after passivation showing degradation of $g_m$ at 200 nS pulse width (d) $f_T$ vs $f_{MAX}$ plot to benchmark our device with other gallium oxide RF device in literature

The sample was passivated with 200 nm thick $Si_3N_4$ deposited by PECVD at 250 $^0$C. The $Si_3N_4$ was removed from the source-drain pad region using $CF_4/O_2$ based ICP-RIE. The cross section schematic of the device is shown in Fig.1 (a) and a FIB cross section image of a fully passivated device is shown in Fig. 1(b)

## III. RESULTS AND DISCUSSION

DC characteristics were measured using 4155B semiconductor parameter analyzer. A 175 nm gate length device gave an on resistance ($R_{on}$) of 23.7 Ω mm at $V_{GS}$ = 5 V (1(c)). Maximum drain current of $I_{DS,MAX}$ = 285 mA/mm and peak $g_m$ of 52 mS/mm were recorded at gate bias of 10 V (Fig 1(d). The peak $I_{DS}$ is comparable to the highest reported current density in gallium oxide FETs [17], [23]. The device shows a depletion mode operation with a threshold voltage of ($V_{th}$) -4 V. The measured on/off ratio is 1.23 x$10^5$ (not shown) can be further increasing the mesa isolation etch to the substrate. The channel sheet resistance was measured to be 14.2 kΩ/□. A mobility of 80 $cm^2$/V.s is extracted from the measured sheet resistance and calculated sheet charge density. TLMs on the regrowth layer give a low 0.078 Ω mm lateral metal/n+ contact resistance with 3.9 x$10^{-7}$ Ω $cm^2$ specific contact resistance. From TLM measurements, channel sheet resistance data and device dimensions, the n++ regrowth/channel interface resistance was estimated to be a high 7.31 Ω mm, contributing to 34% of total on resistance. Atmospheric contaminants at the regrowth interface could be the reason for the high resistance as no pre-treatment was carried out before regrowth. Another plausible reason could be the different growth methods of channel layer (MBE) [28] and n++ regrown layer (MOCVD) [13], [30]. The different methods use different temperature and different growth conditions. Fully MOCVD grown MESFET has been reported with lower interface resistance and contact resistances [5], [22], [31], which shows that lower interface resistances can be achieved.

Three terminal off state Breakdown measurement was performed on the devices using B1505A power device analyzer. We recorded a breakdown voltage of 152 V (Fig. 1 (e)) at $V_{GS}$ = -40 V bias for $L_{GD}$= 355 nm. It corresponds to gate-drain breakdown voltage ($V_{BR}$) = 192 V and 5.4 MV/cm average breakdown field. This $E_{AVG}$ is the highest reported among $\beta$-$Ga_2O_3$ FETs without using any field engineering [32] for sub-micron length devices [33], [34]. (Fig. 1(f))

High frequency small-signal performance was carried out from 100 MHz to 19 GHz using Keysight ENA 5071C Vector Network Analyzer (VNA). The VNA was calibrated using SOLT technique on sapphire calibration standard. Parasitic pad capacitance was de-embedded using an isolated open-pad device structure on the same wafer. Figure 2(a) shows short circuit current gain($h_{21}$), unilateral current gain (U) and maximum available/stable gains (MAG/MSG) at $V_{DS}$= 12 V and $V_{GS}$= 1 V. We extracted current gain cut-off frequency ($f_t$) of $\sim$ 11 GHz and power gain cut off frequency ($f_{MAX}$) of approximately $\sim$ 48 GHz . The peak $f_{MAX}$ value is highest reported in $\beta$-$Ga_2O_3$ FETs.

The expected $f_T$ was calculated from the measured DC $g_m$ and calculated $C_{GS}$ [35]. For $C_{GS}$ calculation, half the channel thickness was used. The measured $f_T$ is significantly lower than calculated values ($\sim$ 25 GHz) Although the device are passivated and current collapse measured in $I_D$-$V_{DS}$ was moderately low (<20 %) (not shown). Pulsed $I_D$-$V_G$ transfer curve shows significant drop of $g_m$ at 200 ns pulse widths (Fig 2 (b)). The source of traps can be attributed to the $SiO_2$ gate dielectric which is deposited *ex-situ* without any surface treatment. Piranha treatment has been reported to reduce traps and hysteresis in transfer curve and capacitance-Voltage curve [36]. This is in contrast to our previously reported HFET [20], [21] where AlGaO layer was deposited *in-situ* during epitaxial device growth.

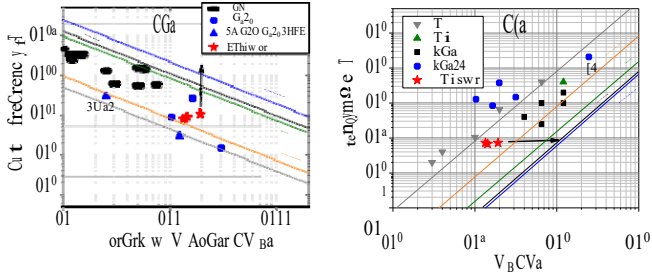

Fig. 3. (a) Johnson's figure of merit showing the trade-off between $f_T$ and maximum operating voltage ($V_{BK}$) or breakdown voltage ($V_{Br}$). (b) Huang's Material Figure of merit ($R_{on}.Q_{GD}$ vs. $V_{Br}$). Materials in the bottom right are considered promising for switching applications

We developed a simplified small signal analytical model (Fig 2 (c)) of the MOSFET using Advanced Design System (ADS). In the model, we used the value of reduced pulsed $g_m$ and calculated $C_{GS}$. The $C_{GD}$ was calculated from the gate-drain separation and measured $R_{DS}$ was used. The $R_i$ was extracted from measured s-parameters at 7 GHz. A good fit is seen between simulated $h_{21}$ and unilateral gain (U) and measured values further confirming the measured RF figures of merit. A scatter plot of $f_T$ vs $f_{MAX}$ previous reported $\beta$-$Ga_2O_3$ devices is shown in (Fig 2 (d)). A $f_{MAX}/f_T$ ratio of 4.3 is obtained because of the T gate structure and record $f_{MAX}$. Similar $f_{MAX}/f_T$ ratio of 3 to 6 [26], [37], [38] have been reported in literature.

Figure 3 (a) shows the trade off between $f_T$ and $V_{br}$ of different materials. With 11 GHz $f_T$ and 192 V breakdown Voltage, we achieved $f_T$. $V_{br}$ product of 2.11 THz.V, which is comparable to mature GaN devices. This makes our device a suitable candidate for L band (1-2 GHz) [39] or S band (2-4 GHz) applications with higher operating voltage ($V_{br}$). Huang's material figure of merit (HMFOM) [40] was calculated by considering that $Q_G$ is dominated by the miller charge $Q_{GD}$ [41]. We estimated $Q_{GD}$ by assuming that full depletion width of the drift region [42], where $Q_{GD}= q.N_D.t_{ch}.L_{GD}.W$. After normalizing with on resistance, we achieved $R_{on}.Q_G$ of 72 m$\Omega$.nC, which surpasses the figure of merit of silicon (Fig. 3(b)) and competitive with SiC and some GaN devices. Similar figure of merit has been reported recently with $Ga_2O_3$ MESFET [43], but with higher $V_{BR}$ coming from large $L_{GD}$ spacing at the expense of high $R_{ON}$.

## IV. CONCLUSION

In conclusion, we have demonstrated a highly scaled $\beta$-$Ga_2O_3$ T-gate MOSFETs, which shows low $R_{ON}$ and high $I_{DS,MAX}$ and transconductance ($g_m$) which are comparable to the state-of-art $\beta$-$Ga_2O_3$ MOSFETS and HFETs. Device shows the highest $E_{AVG}$ for any sub-$\mu$m gate-drain spacing without field plate. We also reported a record $f_{max}$ = 48 GHz among $\beta$-$Ga_2O_3$ FETs. Device surpasses switching figure of merit of silicon at $V_{BR}$ >100 V with low $R_{on}.Q_G$ value. RF performance can be further improved using high-k dielectric and better surface treatment at channel-dielectric interface. This process optimization technique with sub-100 nm gate length paves the way for high-power high frequency applications of future $Ga_2O_3$ MOSFETs